\documentclass{PoS}
\usepackage{amsmath}
\usepackage{braket}
\usepackage{subcaption}
\newcommand{\eps}{\epsilon}
\newcommand{\crefeq}[1]{eq.~(\ref{#1})}
\newcommand{\al}{\alpha}
\newcommand{\cusp}{\gamma_{\text{cusp}}}
\newcommand{\eikJ}{\mathcal{J}}

\title{Wilson-line geometries and the relation between IR singularities of form factors and the large-x limit of DGLAP splitting functions}

\ShortTitle{Relating form factors and PDFs}

\author{\speaker{Calum Milloy}, Giulio Falcioni and Einan Gardi\\
  Higgs Centre for Theoretical Physics, School of Physics and Astronomy \\
  The University of Edinburgh, Edinburgh, EH9 3FD, Scotland, UK\\
        E-mail: \email{calum.milloy@ed.ac.uk}, \email{giulio.falcioni@ed.ac.uk}, \email{einan.gardi@ed.ac.uk}}

\abstract{We discuss the relation between the infrared singularities of on-shell partonic form factors and parton distribution functions (PDFs) near the elastic limit, through their factorisation in terms of Wilson-line correlators. Ultimately we identify the difference between the anomalous dimensions controlling single poles of these two quantities to all loops in terms of the closed parallelogram Wilson loop. To arrive at this result we first use the common hard-collinear behaviour of the two to derive a relation between their respective soft singularities, and then show that the latter is manifested in terms of differing Wilson-line geometries. We perform explicit diagrammatic calculations in configuration space through two loops to verify the relation. More generally, the emerging picture allows us to classify collinear singularities in eikonal quantities depending on whether they are associated with finite (closed) Wilson-line segments or infinite (open) ones.}

\FullConference{14th International Symposium on Radiative Corrections (RADCOR2019)\\ 
9-13 September 2019\\
		Palais des Papes, Avignon, France}

\begin{document}

\section{Introduction}
\label{sec:intro}

Fixed-order calculations in QCD lack predictive power in certain kinematic limits where quantities start to develop large logarithmic enhancements. By factorising these quantities, separating out behaviour that exists at different scales, we can resum these corrections and give all-order results \cite{Sterman:1986aj,Catani:1989ne,Collins:1989bt, FormFactors, Dixon:2008gr}\footnote{see ref. \cite{Falcioni:2019nxk} for a more historical and exhaustive reference list.}.

%Each factor will have anomalous dimensions and these will be related to the anomalous dimension of the original quantity. Some factors take an eikonal form, where partonic fields in the quantity in question are replaced by Wilson lines, see eq.~(\ref{eq:WilsonLineDef}). We shall be exploring two typical configurations of lines in the literature \cite{Korchemsky, Sterman} and also an unfamiliar third \cite{Falcioni:2019nxk}. In doing so we will find that the eikonal forms also satisfy these relations.

As an example of this resummation, the colour singlet on-shell form factor of coloured massless particles with momentum transfer $Q^2$ has the following all-order resummed result \cite{Collins:1989bt, Magnea:1990zb,FormFactors, Dixon:2008gr}
\begin{equation}\label{eq:resumFF}
F(1,\al_s(Q^2),\eps) = \exp\left[\frac{1}{2}\int_0^{Q^2}\frac{d\lambda^2}{\lambda^2}\left(G(1,\al_s(\lambda^2,\eps),\eps) - \cusp\left(\al_s(\lambda^2,\eps)\right)\log\frac{Q^2}{\lambda^2}\right)\right],
\end{equation}
where $\al_s(\lambda^2,\eps)$ is the $d$-dimensional running coupling with $d=4-2\eps$. In the fixed-order calculation, large (double) logarithms in $Q^2/\mu^2$ would develop at large $Q^2$ but in \crefeq{eq:resumFF} these have been resummed. Infrared poles are generated when the integral over $\lambda^2$ is performed. The double infrared poles are governed by the universal spin-independent lightlike cusp anomalous dimension, $\cusp$. The function $G$ has a power expansion in both $\al_s$ and $\eps$ and generates single infrared poles and also finite contributions. We shall define $\gamma_G$ as the piece that gives only the poles,
\begin{equation}\label{eq:gammaGdef}
\int_0^{\mu^2}\frac{d\lambda^2}{\lambda^2}G(1,\al_s(\lambda^2,\eps),\eps) = \int_0^{\mu^2}\frac{d\lambda^2}{\lambda^2}\gamma_G\left(\al_s(\lambda^2,\eps)\right) \,\,\,+\,\,\, \mathcal{O}(\eps^0).
\end{equation}
In contrast, the anomalous dimension $\gamma_G$ depends on whether the particles are spin-1/2 (quarks) or spin-1 (gluons). We shall label it $\gamma_G^{q}$ and $\gamma_G^{g}$ for the two cases respectively.

Another example of resummation occurs in perturbative parton distribution functions (PDFs), $f_{ij}(x)$, the probability of finding parton $i$ with momentum fraction $x$ in parton $j$. After renormalisation they satisfy the renormalisation group (RG) equation \cite{Dokshitzer:1977sg,Gribov:1972ri,Altarelli:1977zs}
\begin{equation}\label{eq:pdfRG}
\frac{d}{d\log\mu}f_{ik}(x,\mu) = 2\sum_j \int_x^1 \frac{dz}{z} \,\,P_{ij}(z)\,\, f_{jk}(x/z,\mu).
\end{equation}
The kernels $P_{ij}$ are called the DGLAP splitting functions. The diagonal splitting functions $P_{qq}$ and $P_{gg}$ diverge in the elastic limit $x\to 1$ as \cite{Korchemsky:1988si,Korchemsky:1992xv,Dokshitzer:2005bf,Belitsky:2008mg}
\begin{equation}\label{eq:Pdiv}
P_{ii}(x) = \frac{\cusp}{(1-x)_+} + B_\delta^i\delta(1-x) + \mathcal{O}((1-x)^0),
\end{equation}
where we see the universal $\cusp$ now as the coefficient of the plus distribution,
\begin{equation}
\int_0^1dx\frac{f(x)}{(1-x)_+} \equiv \int_0^1dx\,\,\frac{f(x)-f(1)}{1-x},
\end{equation}
and the spin-dependent $B_\delta^i$, the coefficient of $\delta(1-x)$.

In both examples soft gluonic radiation dominates at large-$Q^2$ or large-$x$ respectively. In the extreme limit, external particles will then follow the classical trajectory as soft radiation generates no recoil for the emitter. The particles become Wilson lines, defined formally as
\begin{equation}\label{eq:WilsonLineDef}
\phi_n(y,x) = \mathcal{P}\exp\int_x^y dz\,\, n_\mu \,\,A^\mu(z),
\end{equation}
where the Wilson-line is in the direction $n$ between points $x$ and $y$. We will always consider massless particles so the direction will be lightlike, $n^2=0$. The representation of the field $A$ will be the representation of the external particle. Wilson lines are insensitive to the spin of the particle which is why there is no spin-dependence in the leading terms of the form factor and splitting functions. The subleading terms, $\gamma_G$ in the form factor and $B_\delta$ in the splitting functions, have some dependence on hard-collinear regions which are spin-dependent. We shall be considering gauge-invariant correlators of these Wilson lines
\begin{equation}\label{eq:WilsonCorr}
W_{\text{geometry}} = \braket{0|\phi_{n_1}(x_1,y_1)\cdots |0},
\end{equation}
where the geometry depends on the momenta of the external particles. These correlators satisfy RG equations that take the schematic form \cite{Korchemsky,Belitsky:1998tc}
\begin{equation}\label{eq:WilsonRG}
\mu\frac{d}{d\mu}\log W_i =-\cusp \log(m\mu)- \Gamma_i,
\end{equation}
where $m\equiv m(\{n_j\cdot n_k\,,\,x_j\,,\,y_j\,\})$ is a function of the kinematic variables depending on the specific geometry $i$.

\begin{figure}[ht]
  \centering
\begin{subfigure}{0.28\textwidth}\centering
  \includegraphics[width=0.85\textwidth]{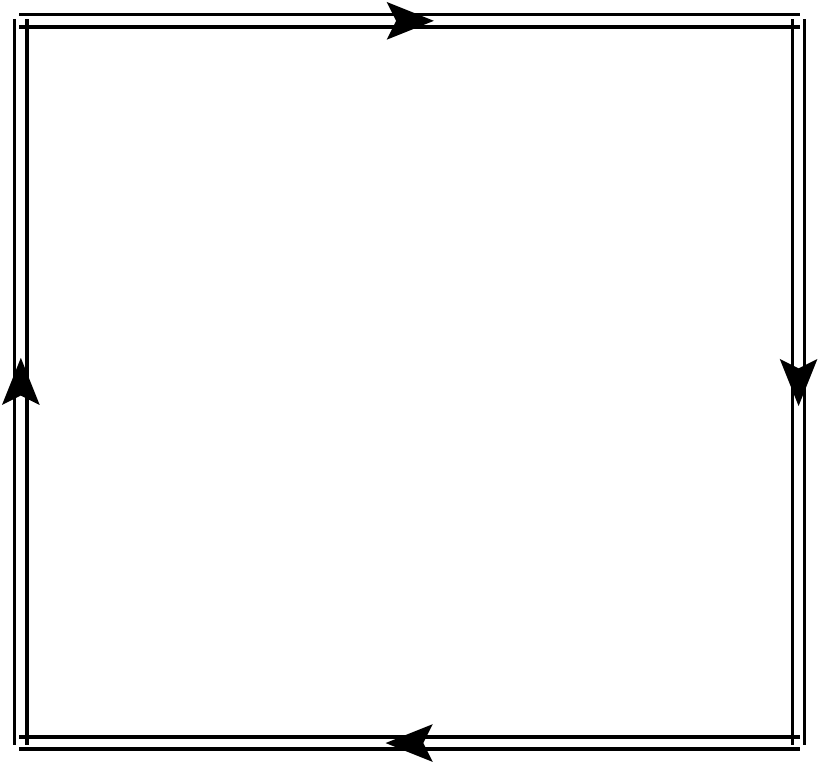}
  \caption{\label{fig:box} The box configuration $W_\Box$}
\end{subfigure}\hskip 4em
  \begin{subfigure}{0.28\textwidth}\centering\captionsetup{width=1.1\linewidth}
  \includegraphics[width=0.85\textwidth]{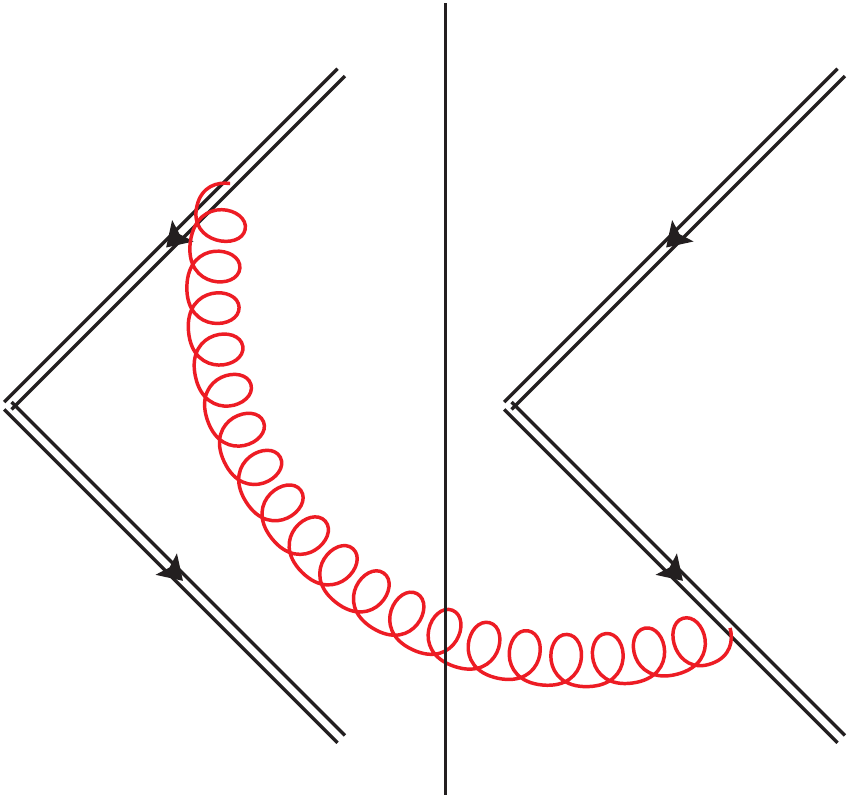}
  \caption{\label{fig:dyWL} A one-loop correction to $W_{\text{DY}}$}
\end{subfigure}
\caption{\label{fig:dyandbox} Two Wilson-line geometries}
\end{figure}

The prototypical configuration is the closed parallelogram ($i=\Box$ in \eqref{eq:WilsonRG}) configuration of Korchemskaya and Korchemsky \cite{Korchemsky}. This geometry is shown in figure.~\ref{fig:box}. There are ultraviolet divergences associated to the cusp but it is infrared finite because the lines are of finite lengths. 

Another configuration is the Drell-Yan soft function shown in figure~\ref{fig:dyWL}. This arises in DY factorisation near threshold and relates quantities defined in \crefeq{eq:gammaGdef} and \crefeq{eq:Pdiv} that states \cite{Sterman:1986aj,Catani:1989ne,Korchemsky:1993uz,Contopanagos:1996nh,Laenen:2005uz}
\begin{equation}\label{eq:DYfactor}
\gamma_G^i - 2B_\delta^i = \Gamma_{\text{DY}}^i/2,
\end{equation}
where $i$ stands for a quark or a gluon. The quantity $\Gamma_{\text{DY}}^i$ is the anomalous dimension defined in \crefeq{eq:WilsonRG}. As a result, it obeys Casimir scaling through three loops,
\begin{equation}\label{eq:Casimir}
\Gamma_{\text{DY}}^g/C_A = \Gamma_{\text{DY}}^q/C_F.
\end{equation}
It is remarkable that \crefeq{eq:DYfactor} gives the difference of two spin-dependent quantities, $\gamma_G$ and $B_\delta$, arising from two different functions, as a spin-independent $\Gamma_{\text{DY}}$ that only depends on the colour representation of the particles. This configuration is actually infrared divergent but safe from ultraviolet singularities.

We make an observation that at two loops, the explicit calculation of $\Gamma_{\text{DY}}$ \cite{Belitsky:1998tc} coincides with that of $\Gamma_\Box$ \cite{Korchemsky}. It is a simple step to then conjecture that it holds to all loops,
\begin{equation}
  \label{eq:DYequalsBox}
  \Gamma_\Box=2\Gamma_{\text{DY}}.
\end{equation}
The identification is not obvious since on the left we have an anomalous dimension associated with finite lines and ultraviolet divergences whereas on the right we have infinite lines and infrared singularities. A natural question arises: can we understand \crefeq{eq:DYequalsBox} through \crefeq{eq:DYfactor} by factorising form factors and PDFs separately?

In this contribution we provide an affirmative answer to the above question and establish that $\gamma_G-2B_\delta = \Gamma_\Box/4$ which are the main results of \cite{Falcioni:2019nxk}. In Section \ref{sec:ff} and \ref{sec:splittingFunctions} we factorise a form factor and a PDF respectively. In Section \ref{sec:relation} we relate the two through their hard-collinear behaviour. In Section \ref{sec:wilsonLine} we explicitly compute the two-loop corrections of the  Wilson-line geometry for the PDF, confirming the relation.

\section{Infrared factorisation of form factors}
\label{sec:ff}
\begin{figure}[ht]
  \centering
  \includegraphics[width=0.5\textwidth]{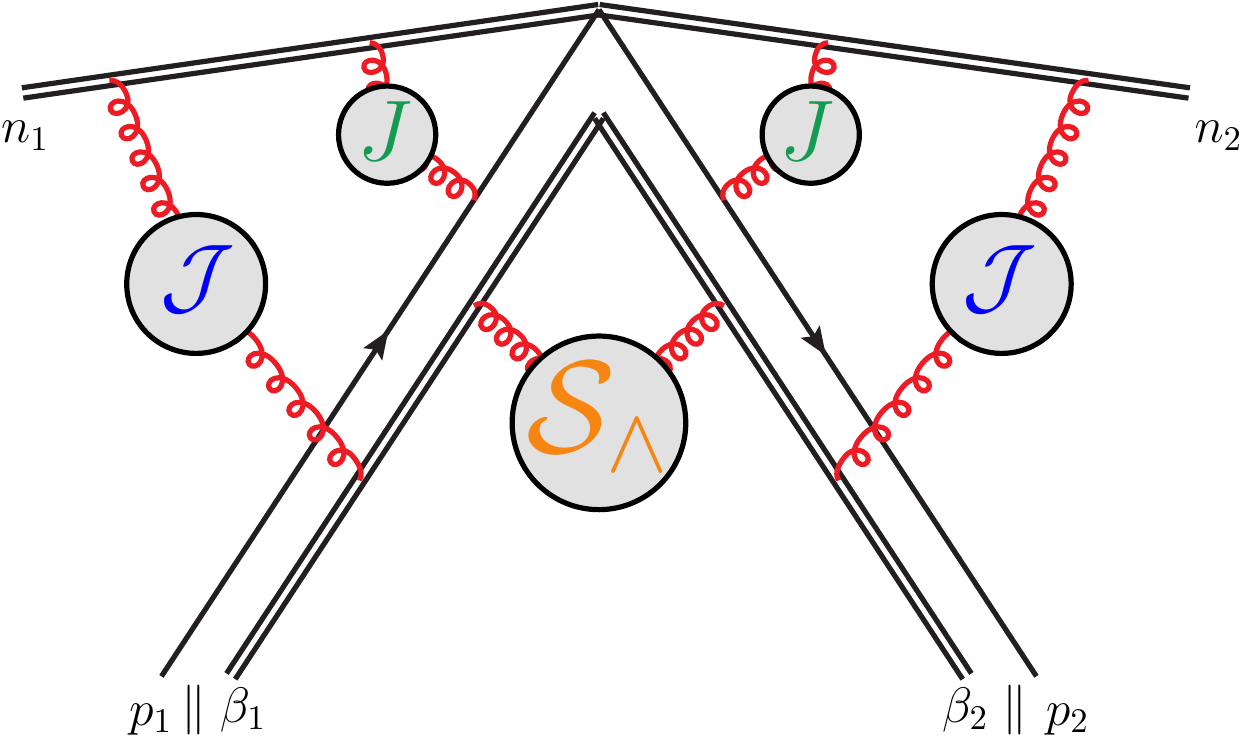}
  \caption{\label{fig:formFactorFactor} Illustration of the infrared factorisation of the form factor}
\end{figure}

We shall start by factorising the form factor in the infrared regime as in refs.~\cite{Collins:1989bt, Dixon:2008gr}. We let the incoming and outgoing momenta be $p_1$ and $p_2$ respectively such that $Q^2=-(p_1-p_2)^2$. Note we shall assume that UV renormalisation of the specific operator used has already taken place. The form factor has two types of IR singularity: collinear and soft. For the collinear region we define jet functions $J_i$ for each external momenta, which for the quark have the form
\begin{equation}\label{jetFunctionDef}
u(p_i)J_i\left(\frac{(2p_i\cdot n_i)^2}{n_i^2\mu^2},\al_s(\mu^2),\eps\right) = \braket{0|T\left[\phi_{n_i}(\infty,0)\psi(0)\right]|p_i},
\end{equation}
where we have defined a non-lightlike auxiliary velocity $n_i$. The soft region is described by two Wilson lines in the wedge ($\wedge$) configuration,
\begin{equation}\label{softFunctionDef}
S_\wedge(\beta_1\cdot\beta_2,\al_s(\mu^2),\eps) = \braket{0|T\left[\phi_{\beta_1}(\infty,0)\phi_{\beta_2}(0,\infty)\right]|0},
\end{equation}
that is $S_\wedge=W_\wedge$ and $p_i=Q\beta_i/\sqrt{2}$. In both functions there is an overlap of the soft-collinear region. This double counting is removed by dividing through by eikonal jets,
\begin{equation}\label{eikJetDef}
\mathcal{J}_i\left(\frac{(2\beta_i\cdot n_i)^2}{n_i^2\mu^2},\al_s(\mu^2),\eps\right) = \braket{0|T\left[\phi_{n_i}(\infty,0)\phi_{\beta_i}(0,\infty)\right]|0}.
\end{equation}
It is easy to check that eqs.~(\ref{jetFunctionDef}), (\ref{softFunctionDef}) and (\ref{eikJetDef}) are all separately gauge invariant. The form factor, $F$, then takes the following form \cite{Collins:1989bt, Dixon:2008gr}
\begin{equation}\label{eq:formFactorFactor}
F = S_\wedge H \frac{J_1}{\eikJ_1}\frac{J_2}{\eikJ_2},
\end{equation}
where a hard function ensures that finite terms match between the left-hand side and right-hand side. We suppressed functional arguments for simplicity. The factorisation is illustrated in figure~\ref{fig:formFactorFactor}.

We shall now isolate the hard-collinear region by considering the ratio of jet functions,
\begin{equation}\label{eq:ratioDef}
J_{i/\eikJ} \equiv \frac{J_{i}|_{\text{pole}}}{\eikJ_i},
\end{equation}
where $J_{i}|_{\text{pole}}$ means only taking the poles of the jet function, since in general it has finite contributions. We shall define the quantity $\gamma_{J/\eikJ}$ to be the single pole anomalous dimension associated with the above ratio.

\begin{figure}[ht]
  \centering
  \begin{subfigure}{0.28\textwidth}\centering
  \includegraphics[width=0.85\textwidth]{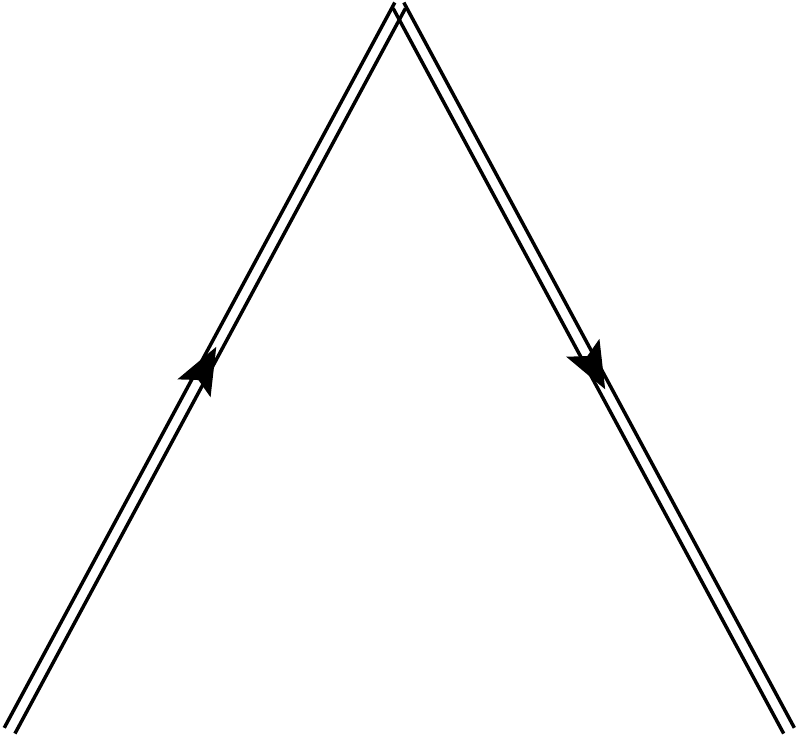}
  \caption{\label{fig:wedge} $W_\wedge$}
\end{subfigure}\hskip 3em
\begin{subfigure}{0.28\textwidth}\centering
  \includegraphics[width=0.85\textwidth]{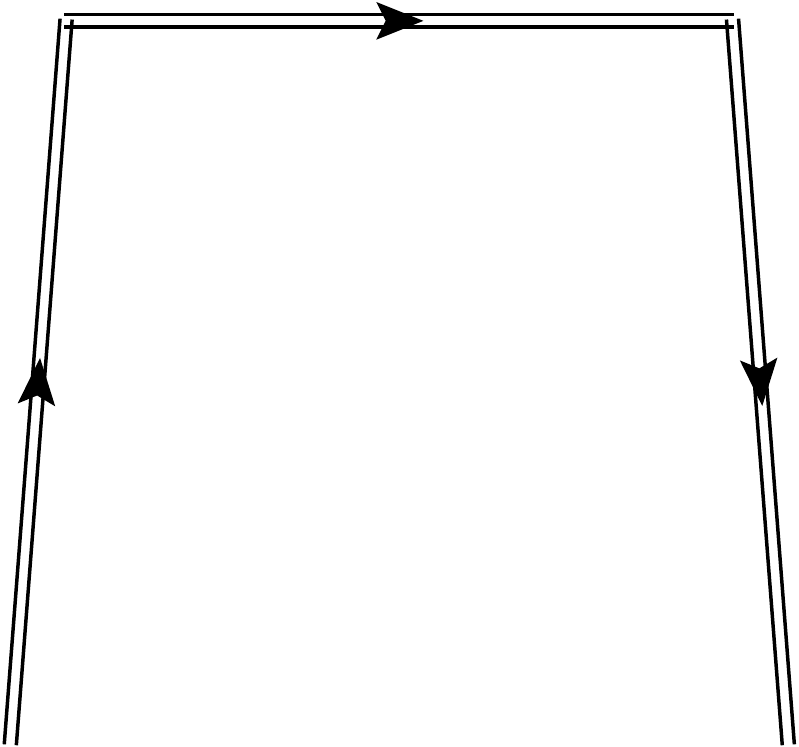}
  \caption{\label{fig:pi} $W_\sqcap$}
\end{subfigure}
\caption{\label{fig:wedgeandPi} Wilson-line geometries associated with the form factor and the PDF respectively}
\end{figure}

By analysing the RG equations of the constituent functions and using eq.~(\ref{eq:formFactorFactor}) we arrive at \cite{Falcioni:2019nxk},
\begin{equation}\label{eq:gammaGRel}
\gamma_G = 2\gamma_{J/\eikJ}-\Gamma_\wedge,
\end{equation}
where $\Gamma_\wedge$ is this the anomalous dimension defined in \crefeq{eq:WilsonRG} with the geometry of Wilson lines shown in figure~\ref{fig:wedge} and known to two loops \cite{Sterman}. Eq.~(\ref{eq:gammaGRel}) could be read as a decomposition of the single pole of the form factor $\gamma_G$ into a spin-dependent hard-collinear term $\gamma_{J/\eikJ}$ and a spin-independent soft term $\Gamma_\wedge$, directly reflecting the single-pole contribution to the factors in eq.~(\ref{eq:formFactorFactor}).

\section{Large-$x$ limit of DGLAP splitting functions}
\label{sec:splittingFunctions}

The light-cone PDF for a quark with longitudinal momentum fraction $x$ in a parton $P$ of momentum $p$ is defined through the following correlator \cite{pdf},
\begin{equation}
f_{qP}^{\text{bare}}(x,\eps) = \frac{1}{2}\int\frac{dy}{2\pi}e^{-iyxp\cdot u}\braket{P|\bar{\psi}_q(yu)\,\gamma\cdot u\,\,\phi_{u}(y,0)\psi_q(0)|P},
\end{equation}
and a similar definition holds for the gluon. The bare PDFs are scaleless and so vanish in dimensional regularisation \cite{Collins:1989gx}. After removal of the UV divergences the renormalised PDFs, $f_{ij}$, satisfy the RG equation in eq.~(\ref{eq:pdfRG}). The renormalised PDFs then have single IR poles. In the following we shall consider the diagonal PDFs $f_{ii}$ and for convenience drop the subscript and only specialise to quarks or gluons when required.

\begin{figure}[ht]
  \centering
  \includegraphics[width=0.5\textwidth]{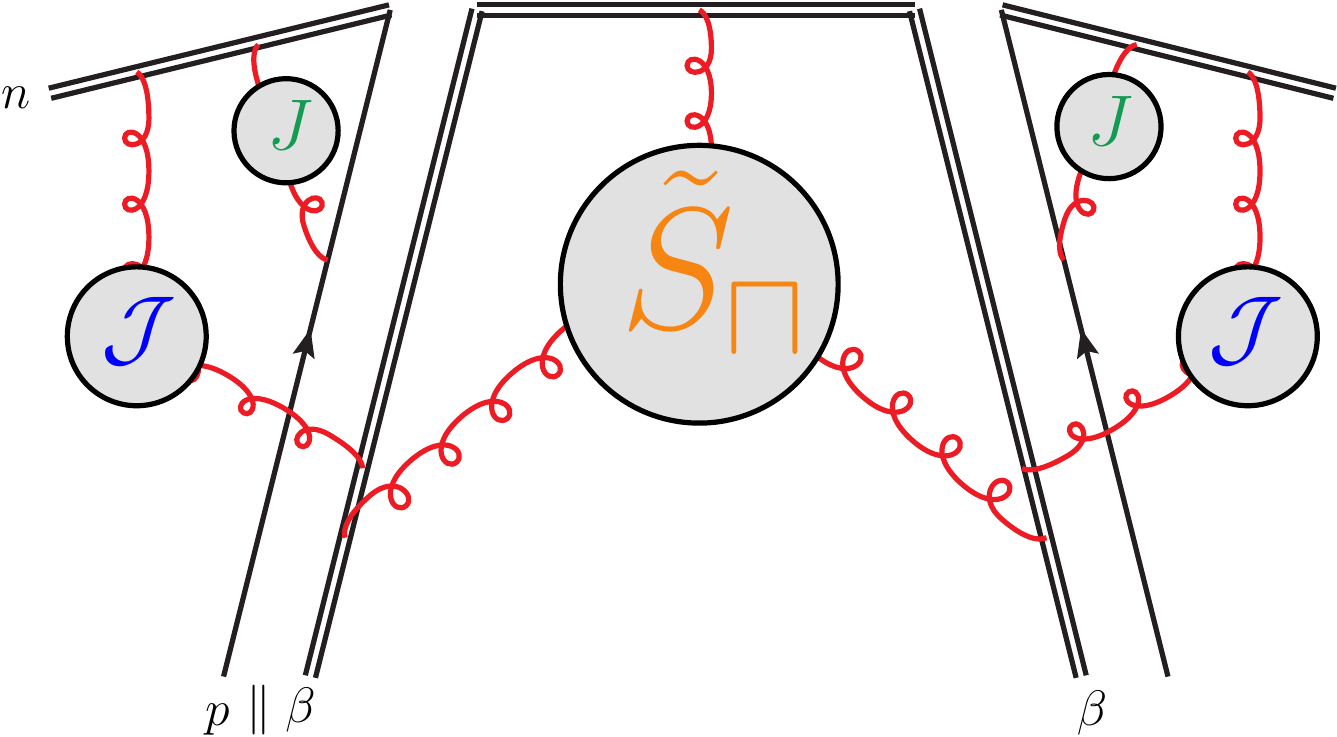}
  \caption{\label{fig:PDFfactor} Illustration of the PDF factorisation at large-$x$}
\end{figure}

Let us now consider the elastic (large-$x$) limit of the PDFs. As $x\to 1$ the external parton loses less and less momentum and so soft gluon emission dominates. This then implies a factorisation \cite{Korchemsky:1988si,Berger:2002sv} in Mellin space,
\begin{equation}
\tilde{f}(N) = \int_0^1 dx\,\, x^{N-1}\,f(x),
\end{equation}
and $x\to 1$ corresponds to $N\to\infty$. The factorisation works much the same way as the form factor. The jet functions are essentially the same because they depend only on the external particle spin and colour and not on the process.%since divergences from the $p$-collinear region will not depend on the Wilson-line in the direction $u$.
Thus they are proportional to $\delta(1-x)$, or $1$ in Mellin space. The difference to the form factor is in the soft function defined as
\begin{equation}\label{eq:SPiDef}
S_\sqcap\left(x,\frac{\beta\cdot u\mu}{p\cdot u},\al_s(\mu^2),\eps\right) = p\cdot u \int\frac{dy}{2\pi}e^{iy(1-x)p\cdot u}W_\sqcap\left(\beta\cdot u\mu,\al_s(\mu^2),\eps\right),
\end{equation}
where $W_\sqcap$ is a Wilson loop in the $\sqcap$ configuration in figure~\ref{fig:pi}. The factorisation is,
\begin{align}
  \tilde{f}(N,\mu) &= H \frac{J_1}{\eikJ_1}\frac{J_2}{\eikJ_2}\tilde{S}_\sqcap(N) + \mathcal{O}\left(\frac{\log N}{N}\right)\nonumber\\
  &= \frac{J_1|_{\text{pole}}}{\eikJ_1}\frac{J_2|_{\text{pole}}}{\eikJ_2}\tilde{S}_\sqcap(N) + \mathcal{O}\left(\frac{\log N}{N}\right),\label{eq:pdfFactor}
\end{align}
where we have used the fact that in a minimal subtraction scheme the $\tilde{f}(N,\mu)$ consist of only poles. The factorisation is illustrated in figure~\ref{fig:PDFfactor}.

The ratio of jet functions in eq.~(\ref{eq:pdfFactor}) is the same as the form factor in eq.~(\ref{eq:ratioDef}), showing that the form factor and the PDF exhibit identical hard-collinear behaviour.  Now considering the RG running of eq.~(\ref{eq:pdfFactor}) we can arrive at \cite{Falcioni:2019nxk},
\begin{equation}
  2B_\delta = 2\gamma_{J/\eikJ} - \Gamma_\sqcap \label{eq:bdeltaRel}
\end{equation}
As in the case of the form factor, we have decomposed the coefficient of $\delta(1-x)$ in the splitting functions into a spin-dependent hard-collinear term $\gamma_{J/\eikJ}$ and a spin-independent soft term $\Gamma_\sqcap$.

\section{The relation}
\label{sec:relation}
Using the universality of the hard-collinear behaviour found in eqs.~(\ref{eq:gammaGRel}) and (\ref{eq:bdeltaRel}) we arrive at,
\begin{equation}\label{eq:theRelation}
\gamma_G - 2B_\delta = \Gamma_\sqcap - \Gamma_\wedge,
\end{equation}
which holds for both quarks and gluons. The terms in the left-hand side of eq.~(\ref{eq:theRelation}) are separately spin dependent while the right-hand side depends only on the colour representation and admits Casimir-scaling (\ref{eq:Casimir}) to three loops (and generalised Casimir-scaling beyond \cite{Moch:2018wjh}). The right-hand side can be read as the ``eikonal'' form of the left.

In \cite{Dixon:2008gr} the difference $\gamma_G-2B_\delta$ was attributed solely to $\Gamma_\wedge$, assuming $B_\delta$ was found from purely virtual contributions to the PDF. Eq.~(\ref{eq:theRelation}) shows that $B_\delta$ has real corrections and this was confirmed by a direct calculation of the diagonal splitting functions at large-$x$ at two loops in \cite{Falcioni:2019nxk}.

We can test eq.~(\ref{eq:theRelation}) to two loops since there are results for $\gamma_G$ \cite{Harlander:2000mg,Ravindran:2004mb,FormFactors}, $B_\delta$ \cite{Floratos:1977au,Floratos:1978ny,GonzalezArroyo:1979df,GonzalezArroyo:1979he,Curci:1980uw,Furmanski:1980cm,Floratos:1981hs,Moch:1999eb} and $\Gamma_\wedge$ \cite{Sterman}. This gives as a prediction for $\Gamma_\sqcap$,
\begin{equation}\label{eq:gammaPiResult}
\Gamma_\sqcap=\left(\frac{\al_s}{\pi}\right)^2\frac{C_i}{2}\left(-2\hat{b}_0\zeta_2-\frac{56}{27}T_fn_f+C_A\left[\frac{202}{27}-4\zeta_3\right]\right)+\mathcal{O}(\al_s^3),
\end{equation}
where $\hat{b}_0=\frac{11}{12}C_A-\frac{1}{3}T_fn_f$ is the coefficient of the one-loop QCD beta function. We now proceed to explicitly calculate $\Gamma_\sqcap$ to two loops in order to compare with the above prediction.
\section{Wilson-line geometries}
\label{sec:wilsonLine}
To find $\Gamma_\sqcap$ we need to calculate the Wilson line correlator $W_\sqcap$. For non-lightlike infinite lines $W_\sqcap$ is known to two loops \cite{Korchemsky:1992xv}. For lightlike infinite lines we encounter scaleless integrals and we need to carefully disentangle UV and IR divergences. It was only recently understood how to achieve this for $W_\wedge$ \cite{Sterman,Erdogan:2014gha,Erdogan:2013bga}. In \cite{Falcioni:2019nxk} we extended this to include a geometry with a finite segment i.e. $W_\sqcap$, which we now review.

We perform the calculation in coordinate space where the gluon propagator is given by,
\begin{equation}
D^{ab}_{\mu\nu}(x) = -\frac{1}{4\pi^{2-\eps}}\frac{\Gamma(1-\eps)}{\left[-x^2+i0\right]^{1-\eps}}g_{\mu\nu}\delta^{ab},
\end{equation}
where the gluon has traversed some space-time separation $x$. $D^{ab}_{\mu\nu}(x)$ is just the $d=4-2\eps$ dimensional Fourier transform of the usual momentum space one. We will utilise non-Abelian exponentiation and compute the logarithm of the Wilson-line correlators in eq.~(\ref{eq:WilsonCorr}) as the sum over so-called \emph{webs} \cite{Sterman:1981jc,Gatheral:1983cz,Frenkel:1984pz,Gardi:2010rn}. In $\log W_\sqcap$ all subdivergences cancel at each perturbative order \cite{Frenkel:1983di,Berger:2002sv,Berger:2003zh,Sterman,Erdogan:2014gha} and the renormalised $\log W_\sqcap$ has the integral form \cite{Falcioni:2019nxk}
\begin{equation}\label{eq:LogWPi}
\log W_\sqcap^{\text{ren}} = -\frac{1}{2}\int_0^{\mu^2}\frac{d\lambda^2}{\lambda^2}\left(2\cusp(\al(\lambda^2,\eps))\log\left(\frac{\rho\mu}{\sqrt{2}}\right) - \Gamma_\sqcap(\al(\lambda^2,\eps))\right),
\end{equation}
where $\rho=i(\beta\cdot uy-i0)$ with $\beta$ and $u$ the directions of the infinite and finite lines resepctively and $y$ is the length of the finite segment.

\begin{figure}[ht]
  \centering
  \includegraphics[width=0.3\textwidth]{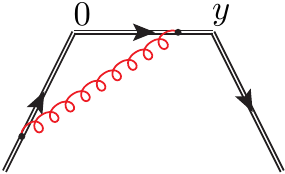}
  \caption{\label{fig:pdfOneLoop} $d_1$\, the one-loop correction to $W_\sqcap$}
\end{figure}

At one-loop there is only one relevant graph\footnote{The ``box''-type graph is proportional to $\beta^2$ and hence vanishes. The Wilson-line self-energy graphs also vanish.} which is shown in figure~\ref{fig:pdfOneLoop}. The Feynman rules give
\begin{equation}\label{eq:d1}
d_1 = \frac{\al_s}{\pi}(\mu^2\pi)^{\eps}u\cdot\beta C_i\Gamma(1-\eps)\int_0^\infty dt\int_0^y ds\,\,(-2\beta\cdot ust+i0)^{-1+\eps}.
\end{equation}
We know that the PDFs vanish for $x>1$. This implies that it is expressible as a function of the variable $\rho=i(\beta\cdot uy-i0)$ \cite{Korchemsky:1992xv}. This ensures there are no singularities in the lower-half $y$-plane allowing us to close the contour in the Fourier transform eq.~(\ref{eq:SPiDef}) and get zero for $x>1$. To proceed in evaluating eq.~(\ref{eq:d1}) we change variables defined through $t=-i\sqrt{2}\lambda$ and $s=-i\sqrt{2}\frac{\sigma}{u\cdot\beta}$. The bare correlator $W_\sqcap$ is $d_1$ and the mirror diagram,
\begin{equation}
\log W_\sqcap^{\text{bare}} = -C_i\int_0^\infty\frac{d\lambda}{\lambda}\int_0^{\rho/\sqrt{2}}\frac{d\sigma}{\sigma}\frac{\al_s\left(\frac{1}{\lambda\sigma}\right)}{\pi}e^{-\eps\gamma_E}\Gamma(1-\eps),
\end{equation}
where we have absorbed variables into the $d$-dimensional running coupling. Now following the prescription in \cite{Sterman} we first expand the integrand in $\eps$,
\begin{equation}
\log W_\sqcap^{\text{bare}} = -C_i\int_0^\infty\frac{d\lambda}{\lambda}\int_0^{\rho/\sqrt{2}}\frac{d\sigma}{\sigma}\frac{\al_s\left(\frac{1}{\lambda\sigma}\right)}{\pi}\left(1+\frac{\zeta_2}{2}\eps^2+\mathcal{O}(\eps^3)\right).
\end{equation}
The first term will develop divergences when $\lambda$ and $\sigma$ both tend to 0, this term will contribute to the cusp. All subsequent terms are of a collinear type when either $\lambda$ or $\sigma$ tend to 0 and contribute to the subleading anomalous dimension $\Gamma_\sqcap$.

For the above result we can drop all terms at $\eps^2$ and above since these will not contribute to the poles. To renormalise we place a short-distance cutoff $1/\mu$ on each integral for the cusp term and just on one integral for the remaining terms. The one loop result is then
\begin{align}
  \log W_\sqcap^{\text{ren}} &=-\frac{C_i}{\pi}\int_{1/\mu}^\infty\frac{d\lambda}{\lambda}\int_{1/\mu}^{\rho/\sqrt{2}}\frac{d\sigma}{\sigma}\al_s\left(\frac{1}{\lambda\sigma}\right)\nonumber\\
  &=\frac{\al_s(\mu^2)}{\pi}\frac{C_i}{\eps}\log\left(\frac{\rho\mu}{\sqrt{2}}\right)\label{eq:oneLoopWPi}
\end{align}
The above is peculiar in that although $W_\sqcap$ has double UV poles it only has a single IR pole. We find that eq.~(\ref{eq:oneLoopWPi}) satisfies eq.~(\ref{eq:LogWPi}) with $\cusp=\frac{\al_s}{\pi}C_i$ and $\Gamma_\sqcap=0$ to one-loop.

\begin{figure}[ht]
  \centering
  \includegraphics[width=1\textwidth]{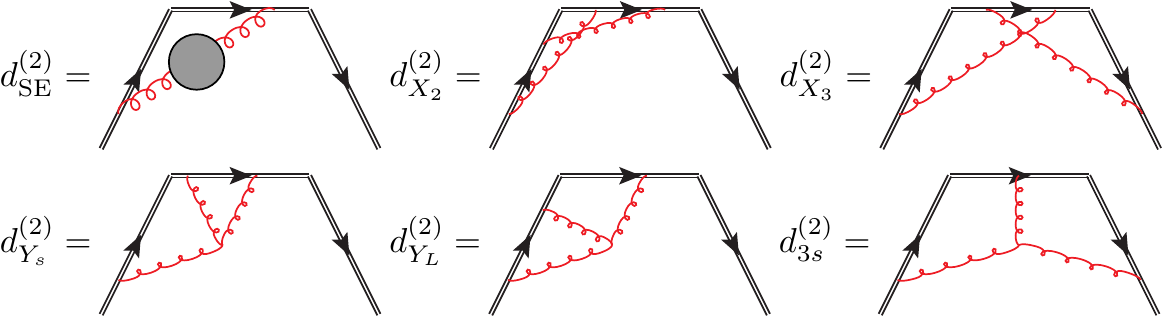}
  \caption{\label{fig:pdfTwoLoop} Two-loop webs for $W_\sqcap$}
\end{figure}

At two loops, because we are calculating $\log W_\sqcap$, we only need the diagrams in figure~\ref{fig:pdfTwoLoop}. The calculations work much the same way as the one-loop case, deriving the same integral representation over one infinite line and a finite one. The procedure is outlined in \cite{Sterman} and details of the calculation of $\log W_\sqcap$ are given in \cite{Falcioni:2019nxk}.
%apart from the graphs that include the three-gluon-vertex, where we have to be careful about whether we first integrate over the position of the vertex or actually evaluating the vertex, since the three-gluon-vertex in coordinate space is a sum of derivatives. By doing the former we are not sensitive to the infinite line behaviour of the Wilson line. It is only when we first integrate by parts using the derivatives from the vertex that we are sensitive to infinite lines \cite{Falcioni:2019nxk,Sterman}.

Summing up all the contributions at two loops and combining with the one-loop result we arrive at,
\begin{align}
\log W_\sqcap^{\text{bare}} &=C_i\int_0^{\infty}\frac{d\lambda}{\lambda}\int_0^{\frac{\rho}{\sqrt{2}}}\frac{d\sigma}{\sigma}\bigg\{\frac{\al_s\left(\frac{1}{\lambda\sigma}\right)}{\pi}e^{-\eps\gamma_E}\Gamma(1-\eps)\left[-1+\frac{\al_s\left(\frac{1}{\lambda\sigma}\right)}{\pi}\frac{11C_A-4T_fn_f}{12\eps}\right]\nonumber\\
  &+\left(\frac{\al_s\left(\frac{1}{\lambda\sigma}\right)}{\pi}e^{-\eps\gamma_E}\right)^2\Bigg[\frac{C_A}{4}\left(\frac{3(-4+3\eps)\Gamma(1-\eps)\Gamma(2-\eps)}{\eps^2(3-8\eps+4\eps^2)}-4\pi\Gamma(-2\eps)\cot\left(\frac{\pi\eps}{2}\right)\right)\nonumber\\
    &-T_fn_f\,\frac{\Gamma(2-\eps)\Gamma(-\eps)}{3-8\eps+4\eps^2}\Bigg]\bigg\}.
\end{align}
To renormalise we perform the same procedure as before: we expand in $\eps$, install ultraviolet cutoffs $1/\mu$ on the integration variables and then integrate which gives,
\begin{align}
\log W_\sqcap&=\al_s(\mu^2)\frac{1}{\eps}\log\left(\frac{\rho\mu}{\sqrt{2}}\right) \nonumber\\&+ \al_s(\mu^2)^2\left\{-\frac{\hat{b}_0}{2\eps^2}\log\left(\frac{\rho\mu}{\sqrt{2}}\right)+\frac{1}{\eps}\left(\frac{1}{4}\Gamma_\sqcap^{(2)}+\frac{1}{2}\gamma_{\text{cusp}}^{(2)}\log\left(\frac{\rho\mu}{\sqrt{2}}\right)\right)\right\},
\end{align}
which satisfies eq.~(\ref{eq:LogWPi}) and where $\Gamma_\sqcap^{(2)}$ agrees precisely with eq.~(\ref{eq:gammaPiResult}), confirming eq.~(\ref{eq:theRelation}).

A more general picture emerges in that divergences in $\log W_\sqcap$ are localised in configuration space and only appear when all vertices in a given web approach a cusp or a line. This is confirmed at two loops by comparing the anomalous dimensions for the different contours,
\begin{subequations}
\begin{align}
 \Gamma_\wedge^{(2)}&=\frac{C_i}{4}\left(-2\hat{b}_0\zeta_2-\frac{56}{27}T_fn_f+C_A\left[\frac{202}{27}\textcolor{red}{-1\zeta_3}\right]\right)\\
 \Gamma_\sqcap^{(2)}&=\frac{C_i}{2}\left(-2\hat{b}_0\zeta_2-\frac{56}{27}T_fn_f+C_A\left[\frac{202}{27}\textcolor{red}{-4\zeta_3}\right]\right)\\
  \Gamma_\Box^{(2)}&=C_i\,\left(-2\hat{b}_0\zeta_2-\frac{56}{27}T_fn_f+C_A\left[\frac{202}{27}\textcolor{red}{-7\zeta_3}\right]\right)
\end{align}
\end{subequations}
All three arise from collinear configurations but are different because of endpoint contributions depending on whether a line is finite or infinite, reflected in the $\zeta_3$ terms. Divergences are blind to the global geometry and allows one to arrive at the general relation,
\begin{equation}\label{eq:PiWedgeBox}
4\left(\Gamma_\sqcap - \Gamma_\wedge\right) = \Gamma_\Box.
\end{equation}
The infinite Wilson line contributions on the left-hand side cancel leaving just one finite Wilson-line segment contribution and $\Gamma_\Box$ is four such finite segments.

\section{Summary}
\label{sec:summary}
In this contribution we factorised a form factor and a large-$x$ PDF separately. We found that they display the same hard-collinear behaviour and only differ in their respective soft regions. Then, arguing that divergences of lightlike Wilson lines in webs are localised to cusps and line segments and are not sensitive to the overall geometry we arrive at the relation,
\begin{equation}\label{eq:endRel}
\boxed{\gamma_G - 2B_\delta = \Gamma_\sqcap - \Gamma_\wedge = \Gamma_\Box/4}
\end{equation}
This links the difference $\gamma_G-2B_\delta$ to $\Gamma_\Box$. It was then checked by an explicit two-loop calculation of $\Gamma_\sqcap$.

Using the DY relation in eq. (\ref{eq:DYfactor}) and eq. (\ref{eq:endRel}), we find $\Gamma_\Box=2\Gamma_{\text{DY}}$. This is an all-orders derivation of the observation made in eq. (\ref{eq:DYequalsBox}). Given that $\Gamma_{\text{DY}}$ can be found from eq. (\ref{eq:DYfactor}) or the actual Wilson-line computation of \cite{Li:2014afw}, the only missing ingredient for a three-loop confirmation is the explicit Wilson-line computation of $\Gamma_\Box$. It also warrants an explanation in terms of the Wilson-line correlators themselves directly relating the two objects shown in figure~\ref{fig:dyandbox}. %It would be interesting to seek a conformal map akin to that found in \cite{Vladimirov:2016dll,Vladimirov:2017ksc} which relates rapidity divergences of parallel lines to UV divergences of semi-infinite lines.

\acknowledgments

GF's and EG's research is supported by the STFC Consolidated Grant ``Particle Physics at the Higgs Centre''; CM's research is supported by a PhD scholarship from the Carnegie Trust for the Universities of Scotland.

\bibliographystyle{JHEP}
\bibliography{proc.bib,bib.bib}

\end{document}